# Rolling Element Bearing Fault Detection and Diagnosis with One-Dimensional Convolutional Neural Network

**Barathan Pubalan[1,2]\*, Muhammad Arif Aiman Jidin[1], Mohd Syahril Ramadhan Mohd Saufi[1,2], Mohd Salman Leong[2], Muhammad Danial bin Abu Hasan[1]**

[1] Faculty of Mechanical Engineering, University Teknologi Malaysia, 81310 Johor Bahru, Johor, Malaysia
[2] Institute of Noise and Vibration, University Teknologi Malaysia, 54100 Kuala Lumpur, Malaysia

*Corresponding Author: barathan@graduate.utm.my

**Abstract:** *Rolling element bearings are critical components in rotating machinery, and their condition significantly influences system performance, reliability, and operational lifespan. Timely and accurate fault detection is essential to prevent unexpected failures and reduce maintenance costs. Traditional diagnostic methods often rely on manual feature extraction and shallow classifiers, which may be inadequate for capturing the complex patterns embedded in raw vibration signals. In this study, a compact one-dimensional convolutional neural network (1D CNN) is developed for automated bearing fault diagnosis using raw time-domain vibration data, eliminating the need for manual feature engineering. The model is trained and evaluated on two established benchmark datasets: the Case Western Reserve University (CWRU) dataset and the Paderborn University (PU) dataset. The CWRU data were segmented based on four distinct motor load conditions (0 HP to 3 HP), with each load scenario trained and tested independently to ensure strict separation and prevent data leakage. The CNN achieved high average test accuracies of 99.14%, 98.85%, 97.42%, and 95.14% for 0 HP, 1 HP, 2 HP, and 3 HP, respectively. On the PU dataset, known for its naturally induced faults and greater operational variability the model achieved a robust average testing accuracy of 95.63%. These results affirm the model's ability to generalize across datasets and varying operating conditions. Further improvements were observed through hyperparameter tuning, particularly window length and training epochs, underscoring the importance of tailored configurations for specific datasets and load conditions. Overall, the proposed method demonstrates the effectiveness and scalability of 1D CNNs for real-time, data-driven bearing fault diagnosis, offering a reliable foundation for condition monitoring in industrial applications.*

## 1. Introduction

The industrial system has made it possible to gather huge amounts of sensor data from a variety of devices. Data-driven technologies are gaining popularity among practitioners to diagnose machine faults (Verdone et al., 2026). The industrial machine usually works in harsh working conditions (e.g high speed, high temperature) continuously which could lead to unexpected breakdowns to machines due to rolling element bearing, gear or other components failure. Therefore, there has always been a need for advanced machine health monitoring techniques to keep industrial machinery functioning correctly and dependably (Fan et al., 2016; Gao et al., 2015; Lei et al., 2014). Rolling element bearings are essential parts of rotational machinery among the many mechanical parts. Smooth-running bearings maintain the efficiency of production processes. Despite the fact that bearings support the shaft, which is subjected to various loading conditions from components such as gears, turbines, and others, they are often overlooked. Bearings typically remain unnoticed until a failure occurs in the plant (Dupont, 2021). The condition of these bearings, including failure points at various locations under different load conditions, significantly affects the performance, stability, and lifespan of the entire machinery system Zhonghai et al. (2018). A widely adopted approach to prevent potential damage is the continuous monitoring of the mechanism's health by analyzing vibration signals. Intelligent fault detection techniques, capable of automatically interpreting these signals and assessing the machine's health, have garnered substantial research interest.

Feature extraction is a critical step in fault diagnosis, as it involves identifying and isolating meaningful patterns or characteristics from raw data, such as vibration signals, to facilitate accurate classification of faults. Traditional feature extraction techniques, such as those based on Empirical Mode Decomposition (EMD), symbolic aggregate approximation, or multifractal analysis, have been widely used in the field of rotating machinery diagnostics. For instance, Asr et al. (2017) developed a feature extraction technique using EMD, where the extracted features were employed to intelligently diagnose faults in rotating mechanisms using a non-naive Bayesian classifier. Similarly, Georgoulas et al. (2015) implemented a symbolic aggregate approximation approach to extract features from bearing signals, which were then classified using a nearest neighbor classifier. Xiong et al. (2016) proposed that vibration signals from bearings exhibit multifractal characteristics, leading to the use of multifractal detrended fluctuation analysis to derive multifractal features for proactive failure diagnostics. Furthermore, Wang et al. (2015) addressed the limitations of the traditional Fourier transform by employing continuous wavelet transform to analyze vibrational signals, with a Support Vector Machine (SVM) classifier used for fault classification.

While these methods have demonstrated effectiveness in extracting features and diagnosing faults, they rely heavily on shallow learning techniques, which have inherent limitations. Shallow learning methods, such as Bayesian classifiers, nearest neighbor classifiers, and SVMs, require manual feature extraction, which is often time-consuming and dependent on domain expertise. The quality of the extracted features directly impacts the performance of the classifier, making the process highly sensitive to the choice of features and prone to human error. Additionally, these methods may struggle to capture complex, non-linear relationships in the data, especially when dealing with high-dimensional or noisy signals. In contrast, modern deep learning approaches, such as convolutional neural networks (CNNs) and recurrent neural networks (RNNs), have gained popularity due to their ability to automatically learn hierarchical features from raw data without the need for manual feature extraction (Zhao et al., 2024). These methods can effectively handle large datasets, capture intricate patterns, and adapt to varying conditions, making them more robust and scalable for fault diagnosis tasks. However, despite their advantages, deep learning models require significant computational resources and large amounts of labeled data for training, which can be a challenge in certain industrial applications.

In summary, while traditional feature extraction techniques have been instrumental in advancing fault diagnosis, their reliance on shallow learning methods introduces limitations in terms of manual effort, sensitivity to feature selection, and inability to model complex data relationships. The shift toward deep learning offers a promising alternative, enabling more efficient and accurate diagnostics by automating the feature extraction process and leveraging the power of advanced machine learning algorithms. Although its application in defect diagnosis is still in the developmental stages, the deep learning approach has demonstrated good results when compared to the traditional machine learning approach (Yang et al., 2020). One of the best methods of deep learning, convolution neural networks (CNN), is also employed to detect bearing problems (Li et al., 2024). The one-dimensional CNN has been used to diagnose real-time motor issues since time-domain signals are the most prevalent data type. It is sometimes possible to employ data that is presented in a two-dimensional manner, like a time-frequency spectrum, to produce images that can be recognized using image processing methods (Ahmed et al., 2024). While many studies rely on feature engineering, this work explores the direct use of raw signals with deep learning for improved automation and accuracy. Hence, this article proposed a fault detection and diagnosis of bearing components with raw vibration data with one-dimensional CNN.

The remainder of this article is organized as follows: Section 2 describes the data collection process, detailing the methods and tools used to gather vibration signals and other relevant data. Section 3 focuses on data processing, including the preprocessing steps and the structure of the 1D CNN model used for fault diagnosis. Section 4 presents the results and discussion, analyzing the performance of the proposed approach and comparing it with existing methods. Finally, Section 5 concludes the article, summarizing the key findings and highlighting potential directions for future research.

## 2. Data Collection

This study utilizes two widely recognized benchmark datasets for bearing fault diagnosis: the Case Western Reserve University (CWRU) dataset and the Paderborn University (PU) dataset, as illustrated in Figures 1 and 2, respectively. Both datasets are frequently used in the research community to evaluate and compare the effectiveness of various fault diagnosis algorithms.

### 2.1 Case Western Reserve University Dataset

The CWRU dataset consists of vibration signals collected from bearings operating under controlled laboratory conditions. Faults of varying sizes (0.007 to 0.021 inches) were artificially introduced at the inner race, outer race, and ball using electro-discharge machining (EDM). The experiments were conducted at different motor loads (0 to 3 horsepower) and speeds (1730 to 1797 rpm). Vibration data were recorded using accelerometers mounted at the drive end and fan end of the motor, with sampling rates of 48,000 samples per second. The controlled environment and well-defined fault types make the CWRU dataset relatively straightforward for machine learning models to classify, providing a clear benchmark for evaluating diagnostic performance.

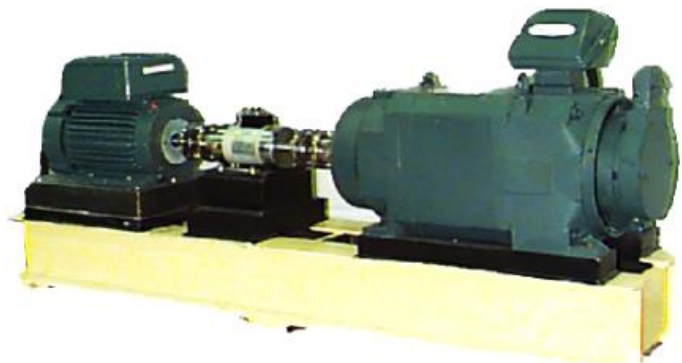

**Figure 1: Bearing experimental setup from CWRU (Smith & Randall, 2015)**

## 2.2 Paderborn University Dataset

The PU dataset, in contrast, presents a more complex and challenging scenario for fault diagnosis. It comprises vibration signals from 32 deep groove ball bearings (type 6203), including both artificially induced and naturally occurring defects. Defects were created using drilling, EDM, and electric engraving, and the dataset covers a wide range of operating conditions with varying rpm, torque, and load. Each signal is sampled at 64 kHz for 4 seconds, resulting in 256,000 data points per record. Only four bearing conditions such as normal, outer race fault, inner race fault, and combined outer and inner race fault are considered, but each is tested under four different operating parameters, resulting in 16 distinct classes. The diversity of fault types, the presence of both artificial and natural defects, and the variability in operating conditions introduce significant noise and non-stationarity, making the PU dataset substantially more difficult for automated classification compared to the CWRU dataset.

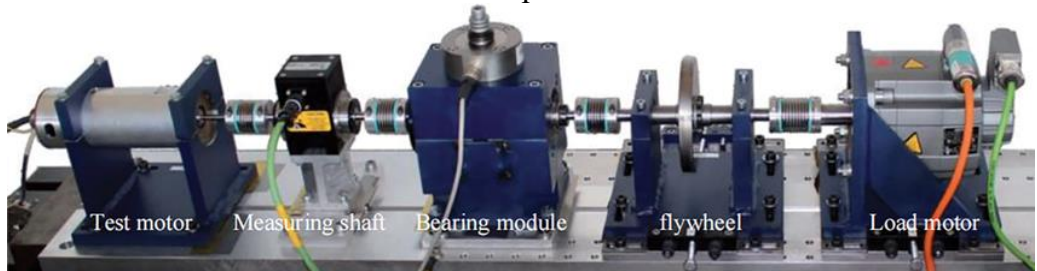

**Figure 2: Bearing experimental setup from PU (Lessmeier et al., 2016)**

## 2.3 Summary

Table 1 and Table 2 summarize the class distributions for the CWRU and PU datasets, respectively. Table 1 presents the structure of the CWRU bearing dataset used in this study. Each row in the table corresponds to a specific bearing condition, such as ball fault, inner race fault, outer race fault, or a healthy bearing. The table details the fault location, the depth of the fault (in inches), and the assigned class label. The dataset was divided into training and testing subsets using a 70:30 split ratio, resulting in 6,481 training samples and 2,778 testing samples. In total, the CWRU dataset comprises 14 distinct classes, each representing a unique combination of fault type and severity, or a healthy state. The balanced distribution of samples across classes and the controlled nature of the faults make this dataset ideal for benchmarking the performance of fault diagnosis algorithms.

Table 2 summarizes the Paderborn University (PU) bearing dataset. Each entry in the table represents a specific bearing under a particular operating condition, with columns indicating the bearing name, fault location (such as healthy, outer race, inner race, or combined faults), and the assigned class label. The dataset consists of a total of 20,435 samples, which were split into 14,304 training samples and 6,131 testing samples based on a 70:30 ratio. The PU dataset includes 16 classes, each reflecting a unique combination of bearing condition and operational setting. Unlike the CWRU dataset, the PU dataset features a greater diversity of fault origins (both artificial and natural), as well as more varied and realistic operating conditions. This complexity is reflected in the larger number of samples and the broader range of class definitions, which together pose a greater challenge for machine learning models and better simulate real-world diagnostic scenarios. For this analysis, each motor load condition (0–3 hp) was treated independently, with its own training and testing partition, enabling a comparative study of diagnostic performance across varying loads.

**Table 1: CWRU bearing dataset data distribution**

| Bearing Condition | Fault Location | Depth of Fault (inch) | Bearing Classification |
|---|---|---|---|

| 14_BA | Ball | 0.014 | Class 1 |
| --- | --- | --- | --- |
| 14_IR | Inner Race | | Class 2 |
| 14_OR1 | Outer Race | | Class 3 |
| 21_BA | Ball | 0.021 | Class 4 |
| 21_IR | Inner Race | | Class 5 |
| 21_OR1 | Outer Race | | Class 6 |
| 21_OR2 | Outer Race | | Class 7 |
| 21_OR3 | Outer Race | | Class 8 |
| 7_BA | Ball | 0.007 | Class 9 |
| 7_IR | Inner Race | | Class 10 |
| 7_OR1 | Outer Race | | Class 11 |
| 7_OR2 | Outer Race | | Class 12 |
| 7_OR3 | Outer Race | | Class 13 |
| N | Healthy | - | Class 14 |

Table 2. PU bearing dataset data distribution

| Bearing Name | Faulty Location | Bearing Classification |
| --- | --- | --- |
| N09_M07_F10_K001_1 | Healthy | Class 1 |
| N09_M07_F10_KA01_1 | Outer Race | Class 2 |
| N09_M07_F10_KB23_1 | Outer Race + Inner Race | Class 3 |
| N09_M07_F10_KI01_1 | Inner Race | Class 4 |
| N15_M01_F10_K001_1 | Healthy | Class 5 |
| N15_M01_F10_KA01_1 | Outer Race | Class 6 |
| N15_M01_F10_KB23_1 | Outer Race + Inner Race | Class 7 |
| N15_M01_F10_KI01_1 | Inner Race | Class 8 |
| N15_M07_F04_K001_1 | Healthy | Class 9 |
| N15_M07_F04_KA01_1 | Outer Race | Class 10 |
| N15_M07_F04_KB23_1 | Outer Race + Inner Race | Class 11 |
| N15_ M07_F04_KI01_1 | Inner Race | Class 12 |
| N15_M07_F10_K001_1 | Healthy | Class 13 |
| N15_M07_F10_KA01_1 | Outer Race | Class 14 |
| N15_M07_F10_KB23_1 | Outer Race + Inner Race | Class 15 |
| N15_ M07_F10_KI01_1 | Inner Race | Class 16 |

## 3. Data Processing and Analysis

The six-stage workflow depicted in Figure 3 begins with the acquisition of vibration records from two benchmark sources such as the Case Western Reserve University (CWRU) Bearing Data Center and the Paderborn University (PU) dataset. Both repositories offer carefully curated, fully labelled signals that span a wide spectrum of defect types and operating conditions, thereby providing an excellent foundation for supervised learning. To convert these continuous records into samples suitable for a 1-D convolutional neural network (CNN), each trace is divided into overlapping windows of fixed length. Specifically, CWRU data are partitioned into 500-point segments with a stride of 300 points, whereas PU signals are trimmed into 1 200-point segments with the stride of 200 points. This controlled overlap preserves temporal continuity, captures transient fault signatures that may straddle window boundaries, and yields a balanced, uniformly sized sample set for subsequent training and validation.

To ensure consistency across conditions, the segmentation procedure was applied separately for each of the four motor load settings in the CWRU dataset. Each subset (0 HP, 1 HP, 2 HP, and 3 HP) underwent an independent training and testing process using the same CNN architecture. This separation was intended to avoid leakage between operating conditions and to assess the classifier's adaptability to different loading scenarios.

Each segmented record (single-axis signal) is fed into a compact 1-D CNN specifically designed for bearing fault diagnosis. The architectural choices were made through empirical experimentation and domain-informed reasoning. The first convolutional layer employs 64 filters with a kernel size of 100 samples to provide a wide receptive field, allowing the network to detect slowly varying amplitude modulations and broadband transient bursts that characterize bearing faults. A smaller kernel would miss these low-frequency envelopes, while larger kernels showed diminishing returns in accuracy and increased computation. The second convolutional layer contains 32 filters with a kernel size of 50, which refines the feature extraction process by focusing on localized, high-frequency fault harmonics derived from the first layer's output. This hierarchical design allows the first layer to act as a coarse-scale envelope detector and the second as a fine-scale feature extractor.

Both convolutional layers use ReLU activation functions, followed by max-pooling (pool size = 4), which reduces computational load while retaining the most salient fault-related activations. Compared with deeper or denser CNNs, this compact two-layer design achieves a favorable trade-off between model size and diagnostic accuracy, making it suitable for edge or embedded applications. The resulting feature map is flattened into a 2816-dimensional vector and passed to a fully connected layer of 100 neurons, which provides sufficient non-linear capacity for class discrimination without overfitting. The final SoftMax layer outputs the probability distribution across fault categories. This architecture, containing fewer than 0.5 million trainable parameters, was deliberately chosen to remain lightweight while maintaining competitive performance relative to other compact CNNs reported in recent literature. Experimental trials with smaller kernels and shallower models resulted in reduced separability in t-SNE visualizations and lower classification accuracy, whereas deeper models offered marginal accuracy gains at the expense of training speed and resource efficiency.

Training and evaluation constitute the final two stages of the pipeline. The network is optimised with the Adam algorithm and categorical cross-entropy loss over 50 epochs, using a mini-batch size of 32. During training the model iteratively adjusts its weights to minimise classification error on the labelled windows, while validation metrics are monitored to detect convergence and guard against over-fitting. After training, performance is assessed on a held-out test set produced by the same segmentation procedure. This rigorous separation of data ensures an unbiased estimate of generalisation capability.

In practice, the architecture routinely converges within a few dozen epochs and delivers near-perfect accuracy across the diverse fault categories present in both datasets confirming its suitability for real-time, high-fidelity condition monitoring of rotating machinery. Additional hyperparameter tuning was conducted for the CWRU dataset across different load levels. While the core CNN architecture remained unchanged, adjustments were made to the learning rate, dropout rate, and batch size during exploratory runs to ensure optimal convergence and generalization in each load case. These variations reflect realistic deployment environments, where load conditions can shift and necessitate dynamic model calibration.

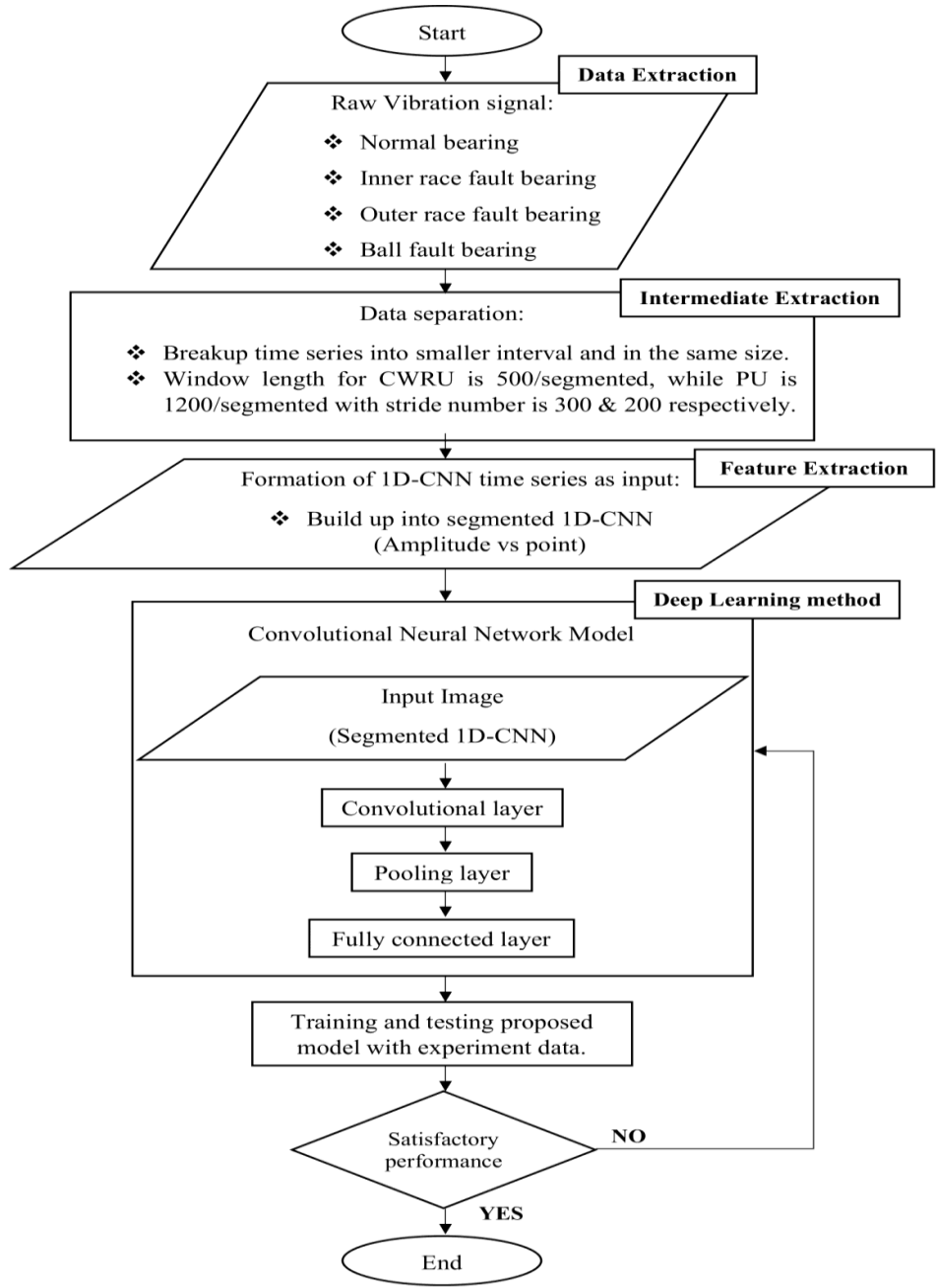

Figure 3: Research Flowchart

## 4. Results and Discussion

### 4.1 Case Western Reserve University Dataset

Confusion matrices and t-distributed Stochastic Neighbor Embedding (t-SNE) visualizations are commonly employed to evaluate model performance and feature separability, respectively. The confusion matrix provides insight into class-wise prediction accuracy and misclassification patterns, while the t-SNE plot visually represents how well the learned features of different fault categories are separated in a low-dimensional space. When the t-SNE visualization shows distinct and compact clusters with minimal overlap between classes, it indicates that the CNN has effectively learned discriminative feature representations.

In contrast, overlapping clusters suggest that the extracted features of certain fault classes are not sufficiently distinguishable, which may lead to potential misclassifications.

The CNN model demonstrates strong capability in classifying various types of machinery defects, as illustrated in Figure 4, which presents the test confusion matrices (left) and t-distributed Stochastic Neighbor Embedding (t-SNE) scatter plots (right) of CNN-extracted features under four different motor load conditions: (a) 0 HP, (b) 1 HP, (c) 2 HP, and (d) 3 HP. Based on the confusion matrices, the test accuracies achieved are 99.14% for 0 HP, 98.85% for 1 HP, 97.42% for 2 HP, and 95.14% for 3 HP. At 0 HP subfigure (a), among the 14 fault classes, several including BDH, BDL, IDH, IDL, N, ODL_2, ODL_3, and ODM achieved perfect classification with 100% accuracy, while the lowest-performing class was BDM with 97%. Under the 3 HP condition (subfigure d), which produced the lowest overall accuracy, classes such as IDL, N, and ODL_1 still maintained 100% accuracy, whereas BDL and ODH_3 recorded the lowest class-wise accuracies, ranging between 85% and 87%. The remaining classes under this condition had accuracies between 95% and 99%.

The corresponding t-SNE scatter plots further illustrate the discriminative power of the CNN model. At lower loads, especially in subfigures (a) and (b), the feature clusters are highly distinct, indicating that the CNN has learned robust and well-separated representations for each fault type. As the motor load increases, however, the clustering quality degrades. In subfigure (d), corresponding to the 3 HP condition, some degree of overlap emerges between fault class clusters, which is consistent with the lower classification accuracy observed. Nevertheless, even at higher loads, the majority of classes remain reasonably well-grouped, and most misclassified samples are located near the correct clusters.

These results suggest that the proposed CNN can extract generalizable features that remain discriminative even under significant load variations. The gradual decline in performance with increasing load reflects the growing complexity of vibration patterns and nonlinear coupling effects in the mechanical system. Despite this, the high classification accuracies and the preservation of cluster boundaries in the t-SNE space indicate that the network has effectively learned invariant representations across operating conditions. This highlights the robustness of the proposed model architecture and its potential applicability in real-world fault diagnosis scenarios, where operating conditions often fluctuate.

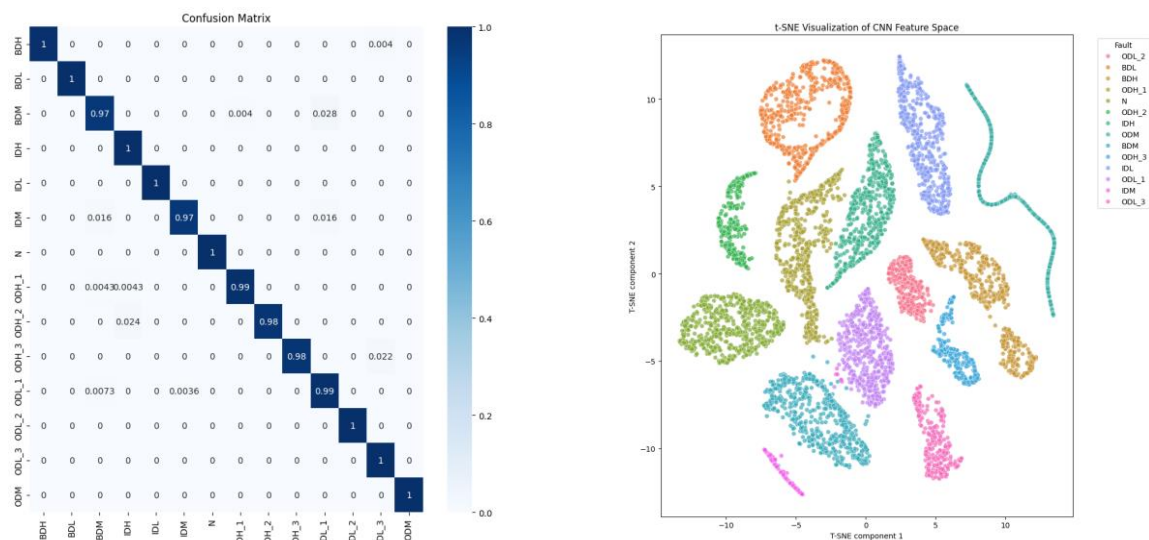

(a) Motor load: 0 HP

(b) Motor load: 1 HP

(c) Motor load: 2 HP

(d) Motor load: 3 HP

**Figure 4: Confusion matrices (left) and scatter plots (right) of the CWRU dataset under different motor load operating conditions**

## 4.2 Paderborn University Dataset

As shown in Figure 5, the CNN model is capable of categorizing various types of machinery faults in the PU dataset, which includes normal bearings, inner race faults, outer race faults, and combined inner–outer race faults across a total of 16 different fault categories. The figure presents confusion matrices (left) and t-SNE scatter plots (right) under different hyperparameter configurations: (a) window length = 500 and stride = 300, (b) window length = 1200 and stride = 300, and (c) early stopping based on validation loss with window length = 1200 and stride = 200. Test accuracy improves significantly across these configurations, starting from 68.19% in (a), increasing to 87.69% in (b), and reaching 95.63% in (c), demonstrating the effectiveness of the hyperparameter tuning process. Notably, the fault classes N09_M07_F10_KA01_1 and N09_M07_F10_KB23_1 achieved perfect classification accuracy (100%) in the best-performing configuration. In contrast, the class N15_M01_F10_K001_1 recorded the lowest classification accuracy at 81%. Although the initial accuracy was relatively low, the tuned model demonstrates substantial performance improvement, confirming that the CNN model, when appropriately optimized, is effective for fault classification in the PU dataset.

This improvement highlights the sensitivity of the CNN to temporal windowing and training stability. Larger window lengths allow the model to capture more complete vibration cycles and fault-related transients, while early stopping prevents overfitting and improves generalization. These results suggest that careful tuning of temporal parameters plays a critical role in enhancing fault recognition performance.

According to the corresponding t-SNE scatter plots (right) under different hyperparameter configurations, the developed CNN model shows increasing effectiveness in distinguishing between normal and faulty data. In subfigure (a), where the model uses a window length of 500 and stride of 300, the t-SNE scatter plot reveals significant overlap between fault classes, indicating limited feature separation. However, as the model is refined through hyperparameter tuning, subfigure (c) which applies early stopping along with a window length of 1200 and stride of 200 shows much clearer and more distinct clustering of classes in the t-SNE space. This suggests that misclassified data are greatly reduced and that the CNN model has learned more discriminative features. The improved cluster separation also reflects the model's ability to map nonlinear vibration dynamics into a feature space that preserves fault-specific characteristics. The strong alignment between confusion-matrix accuracy and t-SNE separability demonstrates that the learned representations are both compact and physically meaningful. Therefore, the constructed 1D Convolutional Neural Network is not only efficient but also highly suitable for diagnosing machinery defects under varying configurations.

It is important to note that the Paderborn University (PU) dataset presents a greater challenge for fault diagnosis compared to the Case Western Reserve University (CWRU) dataset. The PU dataset contains a wider variety of fault types, more complex operating conditions, and a higher degree of signal variability. Unlike the CWRU dataset, which features artificially induced faults under controlled laboratory settings, the PU dataset includes both artificially created and naturally occurring defects, as well as variations in rotational speed, load, and torque. This diversity introduces additional noise and non-stationarity into the vibration signals, making it more difficult for machine learning models to accurately distinguish between fault classes. As a result, while the proposed 1D CNN model achieves near-perfect accuracy on the CWRU dataset, the classification accuracy on the PU dataset is slightly lower. Nevertheless, the model's strong performance on the more complex PU dataset demonstrates its robustness and generalization capability, further validating its effectiveness for real-world bearing fault diagnosis using raw vibration data. These findings collectively indicate that the proposed CNN can adapt to different operating domains without significant degradation in performance. The ability to maintain high accuracy despite higher variability underscores its potential as a general-purpose diagnostic framework suitable for industrial applications.

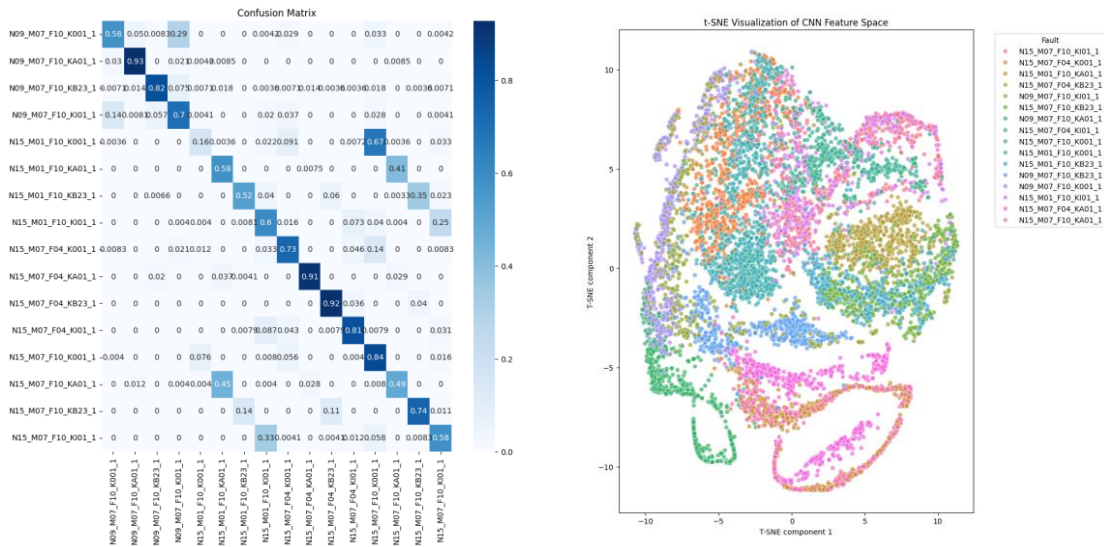

(a) window length = 500 and stride = 300

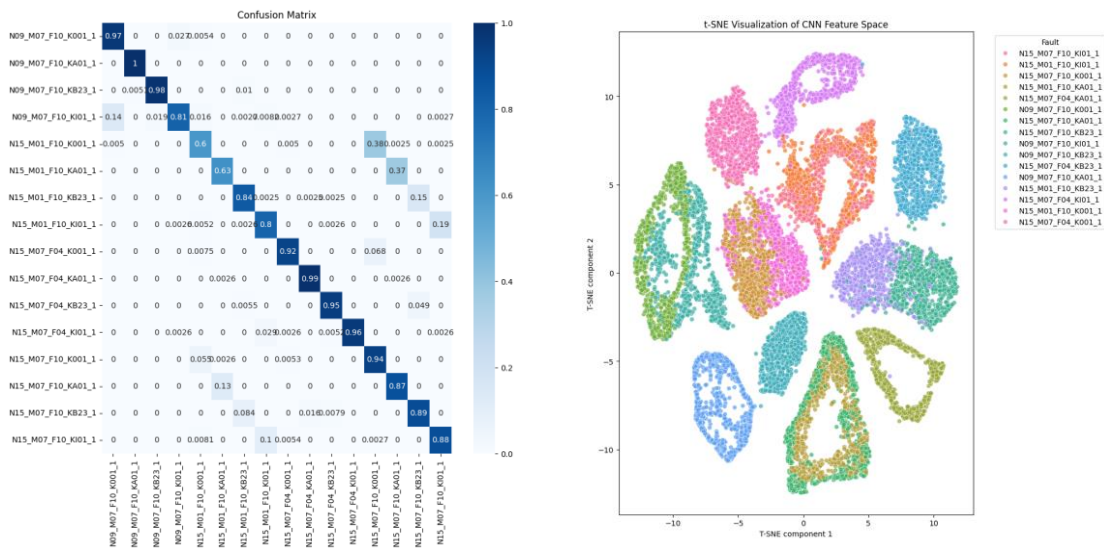

(b) window length = 1200 and stride = 200

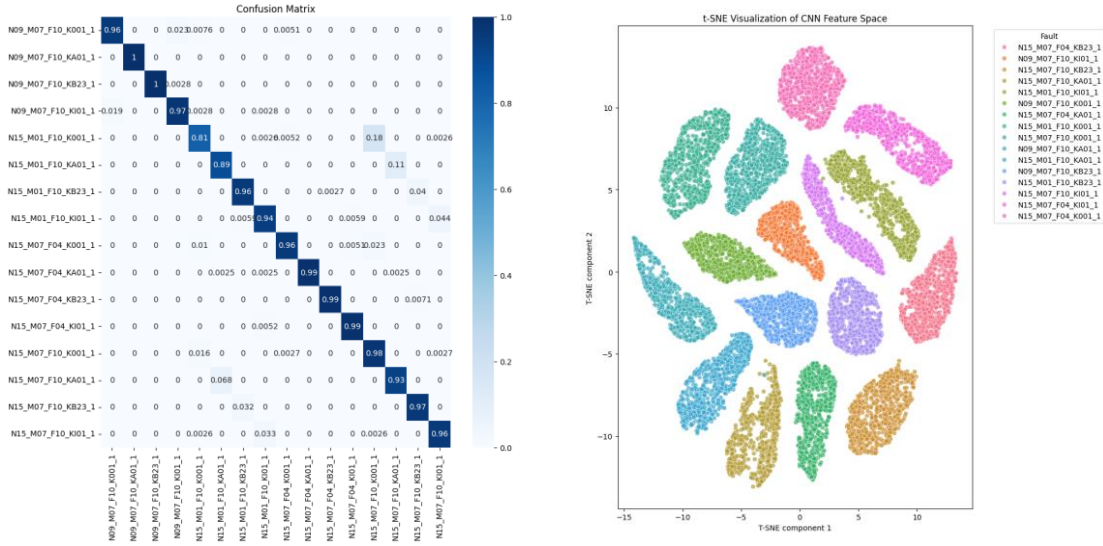

(c) early stopping based on validation loss

**Figure 5: Confusion matrices (left) and scatter plots (right) of the PU dataset under different hyperparameter configurations**

## 4.3 Overall Discussion

The comparative analysis of both the CWRU and PU datasets demonstrates that the proposed 1D CNN framework achieves strong diagnostic performance and reliable feature learning across different operating conditions. The model achieved outstanding classification performance, with average testing accuracies of 97.63% for the CWRU dataset and 95.63% for the PU dataset, clearly demonstrating its effectiveness in diagnosing bearing faults using only raw vibration data without manual feature extraction. While the CWRU dataset yielded near-perfect results due to its controlled laboratory environment, the model also maintained high robustness and adaptability on the more complex PU dataset, where non-stationary signals and load variations introduced greater challenges. This indicates that the CNN effectively captures both local temporal patterns and global fault-related dynamics directly from the raw signals. Furthermore, the strong agreement between the confusion matrices and t-SNE visualizations confirms that the model learns discriminative and well-separated feature representations across domains. Overall, these findings confirm the robustness, generalization capability, and practical potential of the proposed CNN-based framework for real-time, data-driven fault diagnosis in rotating machinery.

## 5. Conclusion

Accurate diagnosis of bearing faults is essential in industrial applications to prevent unexpected downtime and reduce maintenance costs. This study demonstrated the effectiveness of a 1D Convolutional Neural Network (CNN) model using two benchmark datasets: CWRU and PU. The model achieved consistently high classification accuracy across varying motor loads in the CWRU dataset, with testing accuracies of 99.14% at 0 HP and 95.14% at 3 HP, highlighting its robustness under different operating conditions. Similarly, the model achieved a peak accuracy of 95.63% with the PU dataset following careful hyperparameter optimization. Adjustments to hyperparameters such as window length, stride, and early stopping criteria played a crucial role in enhancing model performance, particularly in the more complex PU dataset. These results affirm that CNN models, when properly tuned, are capable of learning discriminative features and accurately diagnosing a wide range of bearing fault types. However, the training of deep learning models remains computationally intensive. Future research should explore strategies to reduce training time and improve efficiency, including the use of high-performance computing resources and lightweight model architectures. Such advancements would support real-time fault detection and broaden the practical applicability of deep learning in industrial maintenance.

However, this study is limited to vibration signals obtained under controlled laboratory settings, and performance may vary under highly non-stationary or real-world industrial conditions. Future research should explore strategies to reduce training time and improve efficiency, including the use of high-performance computing resources and lightweight model architectures. Such advancements would support real-time fault detection and broaden the practical applicability of deep learning in industrial maintenance.

**Acknowledgement**

The work was supported by the Universiti Teknologi Malaysia for supported this work under grant UTM Digital Infuse Research Q.J130000.5124.00L83 and UTM Fundamental Research Grant (UTM-FR) Q.J130000.3824.23H70.